\documentclass[prl,floatfix,showpacs,amsmath,twocolumn]{revtex4}
\usepackage{amssymb}
\usepackage{graphicx}
\usepackage{amsfonts}
\usepackage{amssymb}
\begin{document}

\newcommand{\be}{\begin{eqnarray}}
\newcommand{\ee}{\end{eqnarray}}
\newcommand{\bea}{\begin{eqnarray}}
\newcommand{\eea}{\end{eqnarray}}
\newcommand{\bma}{\begin{subequations}}
\newcommand{\ema}{\end{subequations}}
\def\lR{l^2_{\mathbb{R}}}
\def\RR{{\bf R}}
\def\dd{\delta}
\def\one{{\bf 1}}
\def\kk{{\bf k}}
\def\qq{{\bf q}}
\def\rr{{\bf r}}
\def\nn{{\bf n}}
\def\LL{\textmd{L}}
\def\mm{\textmd{m}}
\def\avw{\overline{\delta \omega}}

\newcommand{\FM}[1]{\bf #1}

\title{Mesoscopic Spin-Boson Models of Trapped Ions}

\author{D. Porras$^1$, F. Marquardt$^2$, J. von Delft$^2$, and J. I. Cirac$^1$}
\affiliation{
$^1$ Max-Planck Institut f\"ur Quantenoptik, Hans-Kopfermann-Str. 1,
Garching, D-85748, Germany. \\
$^2$ Physics Department, Arnold Sommerfeld Center for Theoretical
Physics, and Center for NanoScience,  
Ludwig-Maximilians-Universit\"at M\"unchen, 80333 M\"unchen, Germany.
}

\pacs{PACS}
\date{\today}

\begin{abstract}
Trapped ions arranged in Coulomb crystals provide us with the elements
to study the physics of a single spin coupled to a boson bath. 
In this work we show that optical forces allow us to realize a
variety of spin-boson models, depending on the crystal geometry and
the laser configuration. We study in detail the Ohmic case, which can be
implemented by illuminating a single ion with a travelling wave.
The mesoscopic character of the phonon bath in trapped ions induces new
effects like the appearance of quantum revivals in the spin evolution.
\end{abstract}

\maketitle
%introduction
The problem posed by a two-level system interacting with a
bath of harmonic oscillators, known as the spin-boson model,
appears in condensed matter, atomic physics, and quantum information processing.
It is of fundamental importance, since it represents a paradigm for the study of
quantum dissipation and the quantum-to-classical transition
\cite{Leggett.review,Weiss.book,Weiss.issue}. 
The spin-boson model displays nonperturbative features of
many-body physics, like the localization transition at a critical
dissipation strength. 
For the special case of 
an Ohmic bath spectrum, it may be mapped onto the celebrated Kondo
Hamiltonian. Despite its fundamental importance, experimental
investigations into anything but the weak-coupling regime of the
spin-boson model are still scarce. The localization transition has
been observed in the related Josephson junction systems
\cite{finnish}, while typical solide-state two-level systems feature a
coupling strength much below the critical treshold \cite{japanese}.

Trapped ions provide a clean experimental model system which is ideally suited
for the quantum simulation of condensed matter problems
\cite{spin.models, phonon.models}.
As we show in this work, the spin-boson model 
appears here in a natural way, with a wide range of tunable parameters. 
The role of the two-level system is played by two internal levels of a single ion 
that is part of a Coulomb crystal.
Since ions interact through the Coulomb repulsion, their motion
is described by collective modes with a dispersion relation which
depends both on the direction of vibration 
and on the trapping conditions. 
The internal level of the single ion can be coupled to
the vibrational bath by means of optical forces \cite{Leibfried.review}, 
in such a way that several different spin-boson couplings may be implemented.
Trapped ions offer us a wide range of possibilities for observing the
phenomenology of the spin-boson model, with the advantage that we can
use well-developed experimental techniques for the preparation of both
the initial spin and phonon bath states \cite{Meekhof}. Besides that, the finite
number of vibrational modes forms a mesoscopic environment. This allows us to study
memory effects due to the finite size of the system, such as quantum
revivals in the spin evolution, which are absent in the customary
limit of a macroscopic bath.

%summary/main results

In this work we analyze the spin-boson models which appear in systems
of trapped ions, and we show the following results:
(i) 1D and 2D Coulomb crystals provide us with a variety of
phonon spectral densities, ranging from sub- to super-Ohmic, depending on
the ion crystal dimension and the laser configuration.
(ii) The particular case in which the internal level of  a
single ion is coupled to a Coulomb
chain by a travelling wave corresponds to the Ohmic spin-boson
model.
If the number of ions is large enough, this setup allows us to
tune the interaction strength and recover the standard phenomenology of this
model, such as the quantum phase transition towards localization at strong
coupling.
(iii) For time scales larger than a given revival time, finite size
effects induce the 
reexcitation of the spin after an initial period of relaxation (quantum
revival), which can be observed in a wide range of parameters,
including high temperatures of the phonon bath.
\begin{figure}[h]
\center
\resizebox{\linewidth}{!}{\includegraphics{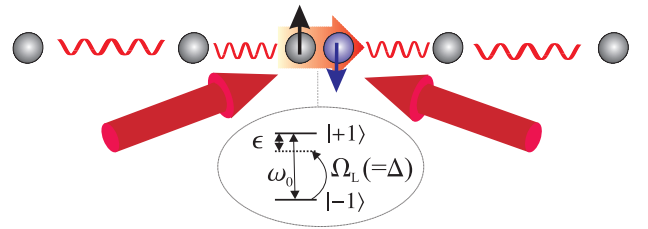}}
\caption{(color online). Scheme of the implementation of spin-boson models
in a Coulomb chain.}
\label{scheme}
\end{figure}

{\it Experimental realization.---}The standard spin-boson Hamiltonian is defined by
\begin{equation}
H = \frac{\epsilon}{2} \sigma_z  
 +  \frac{\Delta}{2} \sigma_x  +  
\sum_n \left( \omega_n a^\dagger_n a_n + 
\frac{\sigma_z}{2} \lambda_n 
( a^\dagger_n + a_n ) \right).
\label{spin.boson}
\end{equation}
$\sigma_\alpha$ are the Pauli matrices, and $\omega_n$ and
$a_n$ are the energies and annihilation operators of the bath phonons,
respectively (we set $\hbar \equiv 1$).
Two hyperfine levels $|-1\rangle$ and $|1\rangle$,
separated by the internal energy $\omega_0$, correspond to 
the eigenstates of $\sigma_z$. 
We propose to experimentally realize the Hamiltonian (\ref{spin.boson}) by
addressing a single ion in a Coulomb crystal in one of two laser
configurations (see Fig. \ref{scheme}):
{\it (i) State dependent dipole force.}
In this setup, a laser induces the tunneling term in Eq. (\ref{spin.boson}), 
in such a way that the tunneling amplitude
$\Delta$ and the bias $\epsilon$ are set by the laser's Rabi
frequency, $\Omega_\LL$, and
the detuning, $\omega_0 - \omega_\LL$, respectively. 
The spin-phonon coupling is induced by subjecting the selected ion to 
an off-resonant standing wave, which creates the state dependent dipole potential 
$V(z) = V_0 \cos^2 \left( k z + \phi \right) \sigma_z$. $k$ and $\phi$
are the wavevector and phase of the standing wave lasers, respectively. 
The operator $z$ is the ion's coordinate in the direction of the
optical force relative to its
equilibrium position. It is readily expressed in terms of phonon operators,
$z = \sum_n {\cal M}_n \bar{z}_n (a_n + a^\dagger_n)$, where 
$\bar{z}_n = 1/\sqrt{2 m \omega_n}$, with $m$ the ion mass, and
${\cal M}_n$ is the amplitude of each vibrational mode $n$ at the ion.
In the Lamb-Dicke regime ($k z \ll 1$) we can expand $V(z)$,
\begin{equation}
H_{\textmd{sw}} = 
\frac{1}{2} \sigma_z \left(V(0) + F z + Q z^2 \right).
\label{dipole.potential}
\end{equation}
$V(0)$ is just a shift of the bias $\epsilon$, and
by choosing $\phi$ appropiately one can retain the linear or
the quadratic terms in (\ref{dipole.potential}) only. 
We focus on the case of a linear coupling ($\phi = \pi/4$, $Q = 0$),
which leads to (\ref{spin.boson}) with $\lambda^\textmd{sw}_n = F {\cal M}_n \bar{z}_n$.
{\it (ii) Polaron coupling induced by a travelling wave.}
In this setup a single laser creates both the spin-boson coupling
and the tunneling term in Eq. (\ref{spin.boson}).
One of the trapped ions in the chain interacts with a travelling wave, 
such that the coupling is given, in a frame rotating with $\omega_\LL$, by
\begin{equation}
H_{\textmd{tw}} = 
\frac{\Omega_\LL}{2} \left( \sigma^\dagger e^{i k z} + \sigma^{-} e^{- i k z} \right).
\label{polaron}
\end{equation}
Hamiltonian (\ref{polaron}) can be recast to fit the expression 
(\ref{spin.boson}) by applying the canonical transformation 
$U = e^{- (i/2) k z \sigma^z}$, which yields
$\lambda^\textmd{tw}_n = k \bar{z}_n {\cal M}_n \omega_n $,
$\Delta = \Omega_\LL$, and $\epsilon = \omega_0 - \omega_L $.

%{\it  situations that we find in 1/2D with different optical forces}

\begin{figure}[h]
\center
\resizebox{3.3in}{!}{\rotatebox{270}{\includegraphics{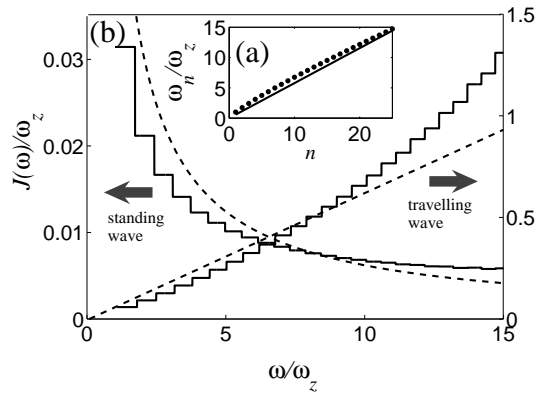}}}
\caption{
Phonon bath properties of a Coulomb chain with $N = 50$ ions.
(a) Dots: Axial vibrational spectrum.
Continuous line: Approximation given by retaining the linear terms in 
$\omega_n^{1D}$. 
(b) Spectral density 
corresponding to spin-boson couplings induced by a standing or a
travelling wave. 
The continuous lines are obtained with the exact
vibrational spectrum, by substituting the delta function in
$J(\omega)$ by a constant function in each interval 
$(\omega_n, \omega_{n+1})$, of height $(\omega_{n+1} -
\omega_{n})^{-1}$. 
The dashed lines are the approximations of Eqs.~(\ref{dos.sw}) and (\ref{dos.tw}).}
\label{vibrational.dos}
\end{figure}

{\it Spectral density.---}The properties of the spin-boson model are determined by the
spectral density, 
$J(\omega) = \pi \sum_{n=1}^N \lambda_n^2 \ \delta(\omega - \omega_n)$.
The model shows a rich phenomenology that has mostly been analyzed
for the case of an infinite gapless bath, such that $J(\omega) \propto \omega^s$. 
In this work we focus on describing experimental conditions 
that lead to this situation.
Two main cases are relevant, namely, 1D and 2D Coulomb crystals. 
Although finite size effects will play an important role, and will
be discussed below, we will first revisit the thermodynamical limit, $N \to \infty$, 
to understand some of the main features.
The axial (1D) or in-plane (2D) modes can be considered approximately gapless, since
the minimum energy is the global trapping frequency along the crystal,
which has to be small enough to guarantee the stability of the latter.
For a 1D Coulomb chain of ions with equally spaced equilibrium
positions, the low energy spectrum of axial vibrations \cite{Morigi}
is approximately given by
$
\omega^{1D}_n =  
v \omega_z \frac{\pi n}{N} \left( 1 - \frac{2}{3} 
\log \left( \frac{\pi n}{N} \right) \right)^{1/2} ,
$
%
%\begin{equation}
%\omega_n = v \frac{2 \pi}{N} \omega_z n + {\cal O}\left( n^2 \log(n) \right)  ,
%\label{1D.dispersion}
%\end{equation}
%
where $v = \sqrt{3 \beta / 2}$, 
$\omega_z$ is the axial trapping frequency, and
$\beta = 2 e^2/(m \omega_z^2 d_0^3)$, with $d_0$ the mean distance.
In order to get a qualitative description,
we will retain the linear term only (see Fig. \ref{vibrational.dos}
(a)), and get the following spectral densities
for each coupling: 
\begin{eqnarray}
J^{\textmd{1D}}_{\textmd{sw}}(\omega) &=& 
(1/v) \bar{{\cal M}}^2 
\left( F z_0 \right)^2 \omega^{-1}, 
\label{dos.sw} \\
J^\textmd{1D}_\textmd{tw}(\omega) &=& 
(1/v) \bar{ \cal{M}}^2
\eta^2 \omega = 2 \pi \alpha \ \omega,
\label{dos.tw}
\end{eqnarray}
where $\bar{\cal M}$ is the mode function of axial phonons at the
addressed ion, $\eta = k / \sqrt{2 m \omega_z}$ is the Lamb-Dicke parameter, and we have
reexpressed $J^{1D}_{\rm tw}(\omega)$ 
as a function of the dimensionless dissipation strength
$\alpha$ \cite{Leggett.review}.
Finite size effects will modify the spectral density of a
Coulomb crystal. Nevertheless, Eqs. (\ref{dos.sw}) and (\ref{dos.tw})
allow us to determine the character of the phonon bath, 
as well as the scaling of its properties with experimental parameters.
This is shown by the comparison with exact numerical calculations in 
Fig.~\ref{vibrational.dos}.
Finally, in a 2D Coulomb crystal, ions arrange themselves in a triangular lattice
\cite{Bollinger}. There, the lowest energy vibrational modes also show a
linear dispersion relation \cite{prl.coulomb.crystals}, 
such that the corresponding spin-boson
models have an algebraic spectral density with exponents 
$s = 0$ \cite{Vojta} and $s = 2$, in the cases of interaction with a
standing wave, or a travelling wave, respectively. 
%

%ion chains: Ohmic/Subohmic, scaling, different transitions
{\it Trapped ion Ohmic model.---}In the following we analyze in detail 
the spin-boson model induced by a
travelling wave interacting with the central ion of a 
Coulomb chain, which leads to (\ref{dos.tw}). 
Since this case corresponds to Ohmic dissipation, it shows a rich 
phase diagram as a function of the
temperature, $T$, and $\alpha$ \cite{note1}.
%Also, since we deal with a mesoscopic environement, the problem is
%related to that of a Kondo impurity in a finite-size metal \cite{Kroha}.
Assume that at $t < 0$ the coupling (\ref{dos.tw}) is off, the phonon
bath is in thermal equilibrium, and the internal state is prepared in 
$| 1 \rangle$ by using standard optical pumping techniques. 
We focus on the evolution of the system at
time $t > 0$, upon suddenly switching on the laser.
In particular we calculate  $P(t) = \langle \sigma^z \rangle$, to
determine  the evolution of the spin under the effect of the tunneling
$\Delta$ \cite{Leggett.review}.
The theoretical description of this system can be addressed within the
Non-Interacting Blip Approximation (NIBA), which consists of a
non-Markovian evolution equation for $P(t)$:
\begin{eqnarray}
\dot{P}(t) &=& \int_0^t  K(t - t') P(t') d t' , \nonumber \\
K(\tau)    &=&  - \Delta^2 {\rm Re} \langle e^{i k z(\tau)} e^{-i k z(0)} \rangle  ,
\label{NIBA}
\end{eqnarray}
%with $Q(\tau) = \langle e^{i k z(t)} e^{-i k z(t')} \rangle$,
where the average is evaluated assuming a thermal state in the bosonic bath.
This expression can be obtained by neglecting spin-bath correlations in
a perturbative expansion up to second order in the tunneling amplitude $\Delta$
\cite{Dekker}. It is well established that it describes correctly the
two limits of weak and strong dissipation, as well as the high
temperature limit.

The finite-size properties of the spin-boson model defined
by Eq.~(\ref{polaron}) can be qualitatively understood by considering the
approximation that vibrational energies are equally spaced by a given
energy $\delta \omega$.  
Under this assumption the Kernel $K(\tau)$ defined by
(\ref{NIBA}) is periodic, and can be written as
$K(\tau) = \sum_n \bar{K}(\tau - \tau_n), \hspace{0.2cm}
\tau_n = n \tau_\textmd{rev}$,
where $\tau_\textmd{rev} = 2 \pi/\delta \omega$ is the vibrational
bath revival time. 
$\bar{K}(\tau)$ becomes equal to the Kernel of the continuous Ohmic
model in the limit $\tau_{\rm rev} \to 0$.
At short times ($t \ll \tau_\textmd{rev}$) the spin
evolution is governed by $\bar{K}(\tau)$, and it shows a similar
behavior as in the case of the continuum spin-boson model. 
The periodic structure of $K(\tau)$ manifests itself at long times 
($\tau \geq \tau_\textmd{rev}$) in the form of quantum revivals in
$P(t)$. 
The spin-boson model of trapped ions satisfies the condition
of equally-spaced vibrational energies only approximately. However
one can define an average energy spacing $\avw$, which
in the limit of large $N$ can be estimated by 
$\avw \approx v \omega_z \pi / N$. 
Thus, in a real Coulomb chain we expect that $\tau_{\textmd{rev}} \approx 2
\pi/ \avw$ defines the time scale which separates the
short-time regime in the evolution of $P(t)$ coinciding with the evolution
for a continuous bath, and the long-time regime displaying the quantum revivals. 

{\it Numerical solution.---}To verify this assumption, 
we obtain the numerical solution of $P(t)$ within
the NIBA, using the exact vibrational modes of a real ion chain.
In order to compare the results obtained in the case of a real Coulomb
chain with those predicted by the ideal Ohmic spin-boson model,
we fit the exact low-energy spectral
density to the form given by (\ref{dos.tw}), and from there we extract
the dissipation strength $\alpha$ which describes the ion vibrational bath.
The NIBA can be justified for a discrete phonon bath as long as the
decay time of $\bar{K}(\tau)$ is much smaller than
$\tau_{\textmd{rev}}$. At short times $t \ll \tau_{\textmd{rev}}$, the
validity of the NIBA in the continuous case implies that the average
in Eq.~(\ref{NIBA}) can be calculated by factorizing the total density
matrix into the spin and phonon reduced density matrices. Due to the
periodic structure of $K(\tau)$, 
brought about by the discreteness of the vibrational
bath, we conclude that $P(t)$ is related to the state of the system at
times $t - n \tau_{\textmd{rev}}$. In the first revival 
($t \approx \tau_{\textmd{rev}}$), $P(t)$ depends on the state of the
system at short-times, where the spin-bath decoupling scheme, which
leads to the NIBA, works. Thus, the first revival is well described
within in the NIBA, and the argument can be easily extended to later revivals.
We checked the NIBA for the finite-size bath vs. the Weisskopf-Wigner
approximation at weak coupling, finding very good agreement.
The numerical solution of the NIBA integro-differential 
equation is performed by means of the numerical method presented in \cite{Wilkie}.
We now discuss in detail the results for two different regimes, 
depending on the phonon temperature.

\begin{figure}[h]
\center
\resizebox{\linewidth}{!}{\includegraphics{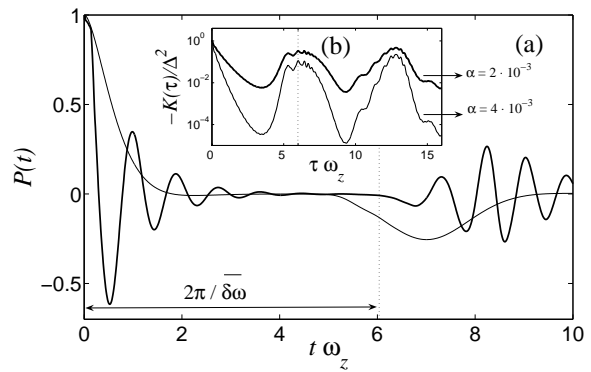}}
\caption{
(a)
Transition from the underdamped
(thick line, $\Delta = 10 \omega_z$, $\alpha = 2 \cdot 10^{-3}$)
to the overdamped regime 
(thin line, $\Delta = 3 \omega_z$, $\alpha = 4 \cdot 10^{-3}$), in the
high-temperature limit of a spin-boson model with $N = 50$ ions, $T
= 250 \omega_z$. 
(b) NIBA time kernels corresponding to these two regimes.} 
\label{revival}
\end{figure}
{\it High temperature regime.---}For sufficiently high temperatures,
the kernel $\bar{K}(\tau)$ decays exponentially with a memory time $\tau_\mm  =
1/(2 \pi \alpha T)$ (we set $k_B = 1$).
In the continuum limit one finds two main
regimes:  $\tau_\mm \Delta > 1 $ (coherent oscillations) 
where $P(t)$ oscillates with $\Delta$ and decays in a time $\tau_\mm$;  
and $\tau_\mm \Delta  < 1$ (overdamped relaxation) where $P(t)$ decays with
a rate $\Gamma = \Delta^2 \tau_\mm$. 
In the case of a finite Coulomb chain, $K(\tau)$ shows an exponential
decay at short times, and additionally, an approximate periodic structure at time
scales $\tau_{\textmd{rev}} \approx 2 \pi / \avw$ (see
Fig. \ref{revival}~(b)). 
Fig. \ref{revival}~(a) shows that 
the behavior of $P(t)$ at short times clearly reveals the transition
between the overdamped and underdamped regimes, as well as the quantum
revivals at $\tau_{\textmd{rev}}$. 
The revival effect can be understood in terms of the 
perturbation of the Coulomb chain created
during the initial spin relaxation, which propagates along the chain,
is reflected at the boundaries, and returns to the selected ion, thus inducing its reexcitation.
%This is a generic effect in finite-size baths. 
%In this one-dimensional setup it is
%particularly pronounced and can be studied in a well-controlled
%and tunable system. 
%If one selects an ion that does not sit at the center of the chain (in contrast
%to what was assumed up to now), propagation times to the left and
%right boundaries are different.
%For such an off-center placement, there will be triple revival events,
%at times 
%$2\tau_{\textmd{rev}}$ and $2 \tau_{\textmd{rev}} \pm 2 \tau '$, where
%$\tau'$ 
%is the time needed for the sound wave to reach the nearer end of the
%chain. 
Interesting geometrical effects on the revivals may also be expected
in $2D$ setups. Revivals in the high-temperature regime could be easily observed in
experiments with trapped ions, since they require neither 
cooling to very low temperatures, nor high values of $\alpha$.

\begin{figure}
\center
\resizebox{\linewidth}{!}{\includegraphics{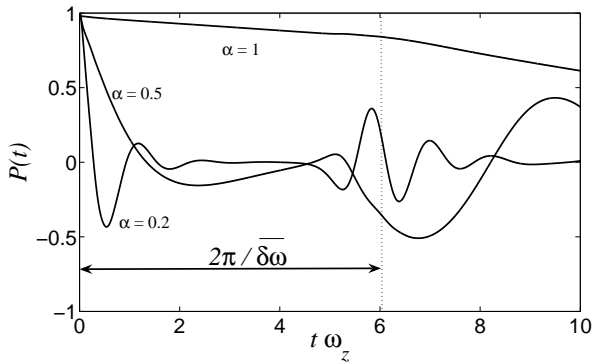}}
\caption{
Evolution of the spin population in the Ohmic trapped ion
spin boson-model, in a Coulomb chain with $N = 50$ ions, $T = 0$, 
$\Delta = 10 \omega_z$, and different dissipation strengths.}
\label{strong.weak}
\end{figure}

\begin{figure}
\center
\resizebox{\linewidth}{!}{\includegraphics{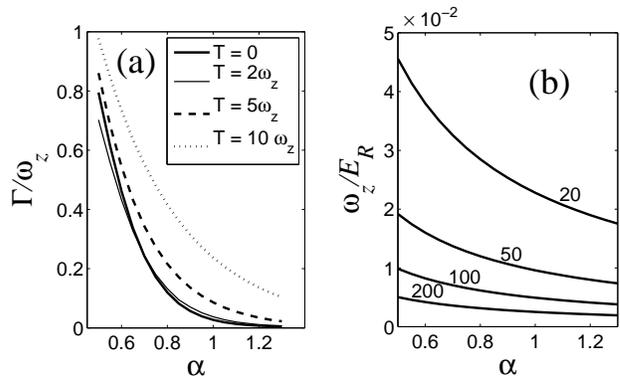}}
\caption{
(a) Evolution of the decay rate in the strong dissipation regime. 
$N = 50$, $\Delta = 10 \omega_z$.
(b) Axial trapping frequency of the trap that is required as
a function of dissipation strength in units of the recoil energy, with
different values of $N$. 
$E_\textmd{R} = k^2/2 m$, with $k$ the wavevector in (\ref{polaron}).
For example,  Be$^+$, $2 \pi / k$ $=$ 300 nm, yields $E_R = 245$ kHz.} 
\label{alpha.evolution}
\end{figure}

{\it Low temperature regime.---}For the Ohmic model in the continuum
and scaling limits,
the evolution of $P(t)$ at $T=0$ is determined by the value of $\alpha$,
in such a way that there are three regimes to
be considered \cite{Leggett.review}: $\alpha < 1/2$ (coherent oscillations),  
$1/2 < \alpha < 1$ (overdamped relaxation), and $\alpha > 1$, in which
case dissipation impedes the decay of $P(t)$ and the system
becomes localized in the initial value of the effective spin
\cite{localization}. 
Since the NIBA 
is known to reproduce the transition between these three regimes 
\cite{Egger}, 
we can use it to investigate whether the Ohmic spin-boson model in
trapped ions shows the same transition. 
Our results are plotted in Fig.~\ref{strong.weak}, where we show that
the relaxation of $P(t)$ shows the same qualitative features as in the
standard Ohmic model, with the appearance of quantum revivals at long times.
The additional localization of the
spin state is clearly evident at values of $\alpha > 1$, although a residual
relaxation process still persists as a consequence of the discreteness
of the bath.
To quantify in more detail the transition to spin localization, we have
calculated the initial decay rates  at short times, as a function of
$\alpha$. Our results in Fig.~\ref{alpha.evolution}(a) show the
slowing down of the spin relaxation with increasing $\alpha$, 
as well as the effect of finite
temperatures. Note that Fig.~\ref{alpha.evolution}(a) corresponds to
the finite-size version of the quantum phase transition to
localization which is found in the thermodynamical limit of the phonon
bath (where $\Gamma$ would vanish above $\alpha = 1$). 

The strong dissipation regime of the mesoscopic Ohmic
spin-boson model requires high values of $\alpha$, and thus of
$\eta^2 \propto1/\omega_z$. 
In the case of a chain with
$N = 50$ ions, the transition to localization can be observed with
$\omega_z$ of a few kHz. 
The axial trapping frequency that is required to realize a model with
a given $\alpha$ decreases with $N$, see Fig. \ref{alpha.evolution}(b).
This condition is difficult to meet in an
experiment, due to the need to cool the Coulomb chain to low
temperatures. However, it must be noted that ground state cooling is
not required. 
Our calculations show that up to temperatures of the order of the axial trapping
frequency the transition to localization can still be observed,
whereas at higher temperatures it is smeared out.

{\it Outlook.---}Let us comment on the possibility of implementing various interesting
experimental situations other than the ones dicussed above. 
In particular, the coupling (\ref{dipole.potential}) allows us to implement 
a bath with $1/f$-noise, a model which is relevant to the description
of decoherence of solid-state qubits (both normal and superconducting)
\cite{Nakamura}.
Besides that, according to our discussion for the case of an off-resonant standing wave
addressing the selected ion, it is possible to tune the spin-boson coupling in such
a way that we implement spin-boson models with couplings quadratic in
the bath coordinate. Again, these are of current concern in the description of
superconducting solid-state qubits operating at the 'sweet spot'\cite{Makhlin},
and could be studied in much more detail in an ion chain model.
Finally, the simultaneous coupling of several
spins to the vibrational bath, by addressing several ions with lasers, would represent
an implementation of a 'many spin-boson' model, where the interplay between the
phonon bath-mediated spin-spin coupling and dissipation could be studied.

Work supported by EU projects (SCALA and CONQUEST),
the DFG through SFB 631, NIM, and an Emmy-Noether grant (F.M.).

\end{document}